\documentclass{article}
\usepackage{spconf,amsmath,graphicx,multirow,hyperref}

\hypersetup{
 colorlinks=true,
 linkcolor=blue,
 filecolor=magenta, 
 urlcolor=cyan,
 pdftitle={Overleaf Example},
 pdfpagemode=FullScreen,
 }

\title{SINGER SEPARATION FOR KARAOKE CONTENT GENERATION}

\name{Hsuan-Yu Lin$^{\dag}$, Xuanjun Chen$^{\ddag}$, Jyh-Shing Roger Jang$^{\star}$}
\address{$^{\dag}$Graduate Institute of Networking and Multimedia \\ 
$^{\ddag}$Graduate Institute of Communication Engineering \\
$^{\star}$Dept. of Computer Science and Information Engineering \\
$^{\dag}$$^{\ddag}$$^{\star}$National Taiwan University, Taipei, Taiwan}

\begin{document}
%
\maketitle 

\renewcommand{\footnotesize}{\scriptsize}

\begin{abstract}
Due to the rapid development of deep learning, we can now successfully separate singing voice from mono audio music. However, this separation can only extract human voices from other musical instruments, which is undesirable for karaoke content generation applications that only require the separation of lead singers. For this karaoke application, we need to separate the music containing male and female duets into two vocals, or extract a single lead vocal from the music containing vocal harmony. For this reason, we propose in this article to use a singer separation system, which generates karaoke content for one or two separated lead singers. In particular, we introduced three models for the singer separation task and designed an automatic model selection scheme to distinguish how many lead singers are in the song. We also collected a large enough data set, MIR-SingerSeparation \footnote{\url{https://gulaerchen.github.io/MIR-SingerSeparation/}}, which has been publicly released to advance the frontier of this research. Our singer separation is most suitable for sentimental ballads and can be directly applied to karaoke content generation. As far as we know, this is the first singer-separation work for real-world karaoke applications.
\end{abstract}

\section{Introduction}
\label{sec:intro}
An audio signal separation model was initially used to separate the target voice from background interference, such as the TasNet model \cite{c5}, which uses the encoder to extract the two-dimensional speech features, and then uses the separation to estimate the speaker mask, and finally uses the decoder to convert the two-dimensional features into a speech waveform to obtain the separated speech. The decoder cannot be perfectly reconstructed, driving in-depth exploration and modification of TasNet, leading to the development of the Multi-phase Gammatone filterbank, which can obtain a better frequency response distribution than random initialization learning.
Compared with a single-channel audio signal, a multi-channel audio signal obtains more spatial information, thereby further assisting speech separation. Wave-U-Net \cite{C13} splices multi-channel signals are input into U-Net and need to change the number of input channels, but the input length of the time domain is usually not fixed if the series is very long, and its resistance to optimization means the traditional RNN model cannot be effectively used. Dual-path recurrent neural networks (DPRNN) optimize RNN in the deep model to process extremely long speech sequences \cite{DPRNN}. Later, a dual-path transformer (DPTNet) \cite{DPTNET} network changed the RNN into a transformer to improve the source separation task.

However, due to the particular complexities of musical structure \cite{drake1993accent}, this specific case of source separation brings challenges and opportunities. A song includes many different instruments and vocals. The instrumental accompaniment and multiple vocal sources cannot be directly separated when applying a multi-channel model. We propose a novel singer separation framework for duet songs, individual or various singers. In addition, we have released four datasets of English and Chinese songs to assist future research in singer separation for duet songs. The proposed method automatically selects an appropriate model for the lyrics language or the number of singers.
The remainder of this paper is organized as follows. In Section 2, we introduce the datasets and their methods of generation. Section 3 introduces the proposed system architecture and auto-selection method. Section 4 presents the content, results, and findings of each experiment. Section 5 presents conclusions and directions for future work.



\section{DATASET}
\label{ssec: DUAL DATASET}

Many datasets have been developed for music source separation. For instance, MUSDB18 is composed of drums, bass, vocals, and remaining accompaniment \cite{musdb18}. The Mixing Secrets dataset is another multi-track dataset used for instrument recognition in polyphonic music \cite{gururani2017mixing}, but features serious leakage between tracks. The Choral Singing Dataset is a good multi-microphone dataset \cite{cuesta2018analysis} that contains the personal recordings of 16 singers from Spain, but has only three songs. Due to the lack of a dataset appropriate for the singer separation task, we have created and released a new dataset consisting of 476 English songs and 500 Chinese songs publicly released on Youtube. The male/female vocalist ratio for the English songs was 269:207, while that for the Chinese songs was 223:277. Based on these ratios, we divided the songs into training and testing subsets (80:20 ratio), ensuring that no vocalists appeared in both subsets. We then converted the music frequency for each song to 8k, used improved accompaniment separation algorithm \cite{master_thesis} to separate the vocal and accompaniment. Finally, we cut the songs into 10-second segments, randomly selecting -5 to 5 SNR to mix pairs of the vocal wave.
The pairs method from left to right in Fig. 1 are the English duet (EN-D), Chinese duet (CH-D), English self-harmonic vocal (EN-S), and accompaniment-vocals-mix (3 channels). In the duet dataset, we extract a vocal segment from different singers and pair them together, mixing the two audio files. Since we hoped that EN-D and CH-D would have similar amounts of data, we repeated the pairing of English singers twice. In the 3 Channels dataset, the vocal part is mixed like the EN-D, and randomly selects an accompaniment segment of the same song to ensure that the vocal data to be analyzed are the same in the EN-D and 3 Channels. The pairing method for EN-S is similar to that for EN-D, except that the former extracts segments for different singers while the latter extracts segments for the same singer. The distribution of pair results are in Table 1.

\begin{figure}
\begin{minipage}[b]{1\linewidth}
 \centering
 \centerline{\includegraphics[width=8.5cm]{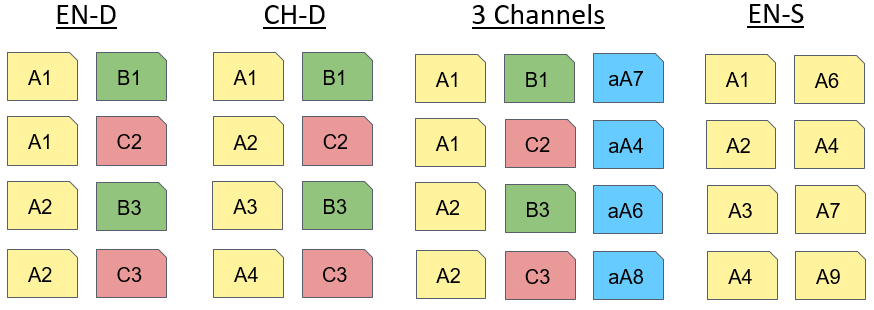}}
\caption{The pairing methods of different datasets. Capital letters represent different singers' vocals (yellow, green, and red blocks). Lowercase a is the accompaniment (blue block). Numbers represent the index of segments in a song.}
\end{minipage}
\end{figure}

\begin{figure}
\centering
\includegraphics[width=8.5cm]{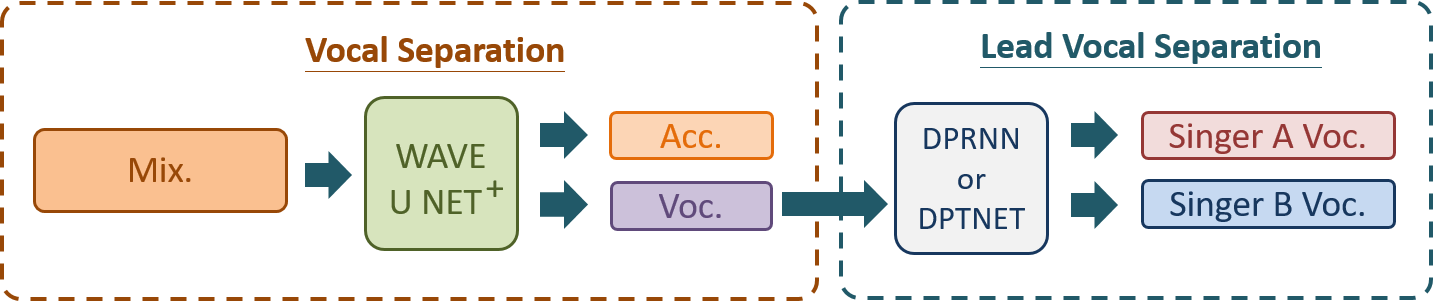}
\caption{System flowchart of singer separation system.}
\label{fig:model}
\end{figure}

\section{SYSTEM FRAMEWORK}
Each song includes many components, such as drums, piano, vocals, harmonies, etc. We divide these components into accompaniment and vocals. Therefore, a singer separation system (SSSYS) includes two stages (Fig. 2), where stage 1 separates the song into accompaniment and vocal, and stage 2 divides the mixed vocals into two vocals.  In the vocal separation stage, we use an improvement of the Wave-U-Net model to accurately differentiate between vocals and accompaniment. We then use the output vocal files as the training materials for the second stage. In the lead vocal separation stage, vocals are mixed to form multiple audio files. We refer to the DPRNN model and DPTNet, converting two-mixed vocal data into singing files of two vocals.

\renewcommand\arraystretch{1.4} 
\begin{table}
\centering
\begin{tabular}{lccccc} 
\hline
\multicolumn{1}{c}{} & \begin{tabular}[c]{@{}c@{}} Train \end{tabular} & \begin{tabular}[c]{@{}c@{}} Valid \end{tabular} & \begin{tabular}[c]{@{}c@{}} Test \end{tabular} & \begin{tabular}[c]{@{}c@{}}Duration\end{tabular} \\ 
\hline
\begin{tabular}[c]{@{}c@{}} EN-D \end{tabular} & \begin{tabular}[c]{@{}c@{}}21772 \end{tabular} & 11373 & 12970 & 384hr 17.5min \\ 
\begin{tabular}[c]{@{}c@{}} CH-D \end{tabular} & \begin{tabular}[c]{@{}c@{}}16495 \end{tabular} & 4059 & 4059 & 205hr 6.5min \\
\begin{tabular}[c]{@{}c@{}} EN-S \end{tabular} & \begin{tabular}[c]{@{}c@{}}17804 \end{tabular} & 6489 & 6489 & 256hr 31min\\
\begin{tabular}[c]{@{}c@{}} 3 Channels \end{tabular} & \begin{tabular}[c]{@{}c@{}} 21772 \end{tabular} & 11373 & 12970 & 384hr 17.5min \\
\hline
\end{tabular}
\caption{Data distribution of each dataset.} 
\label{fig:Data distribution}
\end{table}

\subsection{Vocal separation in Wave-U-Net$^+$}
\label{ssec: VU}
U-Net was first proposed by Ronneberger in 2015 \cite{ronneberger2015u} and used to segment biomedical images. The model can be divided into three parts: encoder, decoder, and skip-connection. Two years later, Jansson proposed using U-Net in song vocal separation by spectrum \cite{jansson2017singing}. The main difference between Jansson's U-Net and Ronneberger's original U-Net is that each encoder layer uses two strides and no pooling layer, which is closer to the model proposed by S. Boll \cite{boll1979suppression}. Combining the above methods, there are three characteristics in the Wave-U-Net$^+$ model \cite{master_thesis}:
\begin{itemize}
 \item [1)]Use the encoder and decoder architectures in Ronneberger's U-Net.
 \item [2)]Adjust the stride setting of the pooling layer so that after the input spectrum passes through the pooling layer, the time dimension remains unchanged. 
 \item [3)]Refer to Jansson's U-Net and set the number of model encoders to six layers.
\end{itemize}

The Wave-U-Net$^+$ model is used in the first stage. The input is music, and the outputs are accompaniment and vocals.


\begin{figure}
\begin{minipage}[b]{1\linewidth}
  \centering
  \centerline{\includegraphics[width=8.5cm]{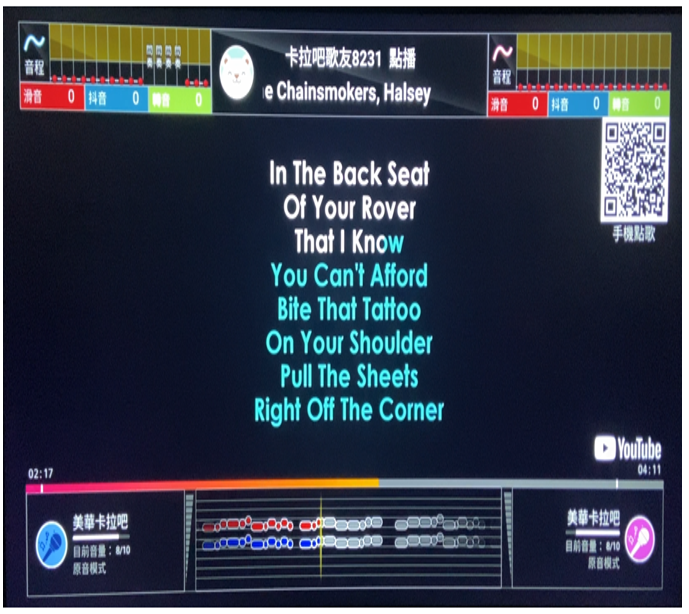}}
  \centerline{(a) The user interface of the karaoke application.}\medskip
\end{minipage}
\hfill
\begin{minipage}[b]{1\linewidth}
  \centering
  \centerline{\includegraphics[width=8.5cm]{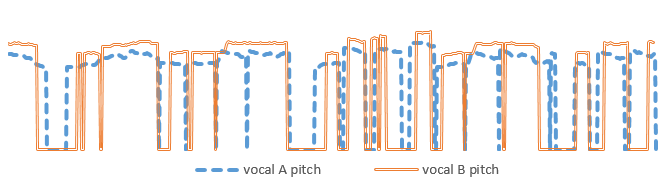}}
  \centerline{(b) Pitches data of actual singing.}\medskip
\end{minipage}
\caption{The pitches trend in a song, blue and red lines at the bottom of (a) are the pitches of the two lead singers, and they correspond to vocal A and B of (b), respectively. }
\label{fig:speech_sing}
\end{figure}

\subsection{Lead vocal separation in DPRNN or DPTNet}
\label{ssec: Stage 2: Singer Separation}
Dual-path RNN (DPRNN) consists of three stages: segmentation, block processing, and overlap \cite{DPRNN}. It divides longer audio inputs into smaller chunks. It iteratively applies intra-inter-chunk operations, where the input length (chunk size) can be proportional to the square root of the original audio length in each operation. Dual-path transformer Network (DPTNet) also has three steps: encoder, separation layer, decoder \cite{DPTNET}. By introducing an improved transformer, the elements in the speech sequence can interact directly, thus enabling DPTNet to model speech sequences through direct context-awareness.

In the second stage, we use the DPRNN or DPTNet model, split the mixed-vocals into two vocals.


\subsection{Model auto-selection}
\label{ssec:Model architecture}
Given an input song, we can’t choose the corresponding model because we don’t know the language and whether the same singer performs the harmony vocals. We found the pitch trends of the two singers are almost the same in most songs (Fig. 3), so if the two-channel pitch trend after separation is the same, we can consider this model to have the best separation effect. Therefore, the automatic model selection method is:

\begin{equation}
	\mathop{\arg\min}_{m}
 \sum_{i}{\|v_{mAi}-v_{mBi}\|}
\end{equation}

\noindent
where $m$ is the model type (EN-Duet, CH-Duet, and EN-Self), $A$ and $B$ are the vocals A and B, and $v$ is the difference of pitches in same vocal:

\begin{equation}
\begin{cases}
v_{mAi} = p_{mAi+1}-p_{mAi} & \\
v_{mBi} = p_{mBi+1}-p_{mBi} & 
\end{cases}
\end{equation}


\noindent
where $p$ are the pitches obtained from CREPE \cite{crepe}. It is worth noting that using the sliding window with a unit of three ensures the algorithm skips audio segments when only one singer is singing (3). When the pitches of one vocal are all zero, we set a high penalty value, making the system select the best separation model.

\begin{equation}
v_{mAi} - v_{mBi} =
\begin{cases}
v_{mAi} - v_{mBi}
&
\text{, when $p_{i-1}$, $p_{i}$ and $p_{i+1}$ $>$0 }
\\
0& \text{, otherwise}
\end{cases}
\end{equation}

Through the auto-selection algorithm (1), we obtain a separation result with the closest trend as the final system output.

\section{EXPERIMENTS AND RESULTS}
\label{sec:experimental}
In all experiments, we evaluate system effectiveness based on signal fidelity. The degree of improvement in signal fidelity can be measured by SI-SNR improvement (SI-SNRi) \cite{sisnr}, and signal-to-distortion ratio improvement (SDRi) \cite{sdr}, which are both often used in audio source separation systems. We calculate these two indicators for the vocal part only, thus increasing the fairness of the experimental results.

\renewcommand\arraystretch{1.4} 
\begin{table}
\centering
\begin{tabular}{lccc@{}}
\hline
& \multicolumn{3}{c}{\bf Experimental results} \\ 
\cline{2-4} 
& \begin{tabular}[c]{@{}c@{}}Training \\Time\end{tabular} &\begin{tabular}[c]{@{}c@{}}SI-SNRi\\(dB)\end{tabular}&\begin{tabular}[c]{@{}c@{}}SDRi\\(dB)\end{tabular}\\
\hline
\begin{tabular}[c]{@{}c@{}}DPRNN \\ 3 channels\end{tabular}& 9D3H48M & 3.2412 & 4.0397\\
\hline
\begin{tabular}[c]{@{}c@{}}SSSYS \\ (DPRNN)\end{tabular}&10D23H46M&8.2679&8.7844\\
\hline
\begin{tabular}[c]{@{}c@{}}SSSYS \\ (DPTNet)\end{tabular}&11D21H50M&
\textbf{9.3741}&
\textbf{8.8861}\\
\hline
\end{tabular}
\caption{Comparison of different model on English duet data.} 
\label{table: compare3}
\end{table}

\renewcommand\arraystretch{1.4}
\begin{table*}
\setlength{\tabcolsep}{4.7mm}{
\centering
\begin{tabular}{lccccccc} 

\hline & \multicolumn{1}{l}{} & \multicolumn{2}{c}{\begin{tabular}[c]{@{}c@{}}EN-D \\Dataset\end{tabular}} & \multicolumn{2}{c}{\begin{tabular}[c]{@{}c@{}}CH-D
\\Dataset \end{tabular}} & 
\multicolumn{2}{c}{\begin{tabular}[c]{@{}c@{}}EN-S
\\Dataset \end{tabular}} 
\\ 
\cline{3-8}
\multicolumn{1}{c}{} & \begin{tabular}[c]{@{}c@{}}Training\\Time\end{tabular} & \begin{tabular}[c]{@{}c@{}}SI-SNRi\\(dB)\end{tabular} & \begin{tabular}[c]{@{}c@{}}SDRi\\(dB)\end{tabular} &
\begin{tabular}[c]{@{}c@{}}SI-SNRi\\(dB)\end{tabular} & \begin{tabular}[c]{@{}c@{}}SDRi\\(dB)\end{tabular} &
\begin{tabular}[c]{@{}c@{}}SI-SNRi\\(dB)\end{tabular} & \begin{tabular}[c]{@{}c@{}}SDRi\\(dB)\end{tabular}

\\ 
\hline
\begin{tabular}[c]{@{}c@{}}EN-Duet\\Model\end{tabular} & \begin{tabular}[c]{@{}c@{}}10D\\23H46M\end{tabular} & \textbf{8.2679} & \textbf{8.7844} &
10.0857 & 10.4660 &
4.5389 & 3.2236 \\
\hline

\begin{tabular}[c]{@{}c@{}}CH-Duet\\Model\end{tabular} & \begin{tabular}[c]{@{}c@{}}6D\\1H59M\end{tabular} & 
6.7841 & 7.3587 & 
\textbf{10.8926} & \textbf{11.1288} & 
- & - \\
\hline

\begin{tabular}[c]{@{}c@{}}EN-Self\\Model\end{tabular} & \begin{tabular}[c]{@{}c@{}}7D\\7H12M\end{tabular} & 
6.6387 & 7.1639 & 
- & - & 
\textbf{4.7366} & \textbf{5.2674} \\
\hline

\end{tabular}}

\caption{Comparison results in English duet, Chinese duet, and English self harmonic dataset.} 
\label{fig:result2 table}

\end{table*}


\subsection{Experimental configurations}
\label{ssec:Experiment}

Experiments were run on a computer running the Ubuntu 18.04.5 LTS" operating system, with 8 AuthenticAMD CPUs, a GeForce RTX3090 GPU, 64GB RAM, and a 1TB SSD. Each model was trained for 100 epochs, with a learning rate of $5e^{-4}$, and using Adam as the optimizer. A stop condition is set if no best model is found in the validation set for ten consecutive epochs. As a baseline, the input is the 3 Channels dataset, and we use DPRNN 3 channels to directly separate the accompaniment and vocals of two different singers. The hyperparameters are the same as the SSSYS (DPRNN).

\subsection{Singer Separation Results}
\label{ssec:Results}

We first compare the DPRNN 3 channels and the SSSYS with the DPRNN or DPTNet in the second stage (Table 2). DPRNN 3 channels only obtains evaluation results of 3.2412 and 4.0397, while SSSYS exceeds 8 in SI-SNRi and SDRi, thus indicating our proposed system obtains better results.

Other languages use other models in speech separation, so we also divide songs into various languages for comparison of singer separation (Table 3). The performance of the English duet model in the English duet dataset is 8.2679 and 8.7844, outperforming Chinese duet model (6.7841 and 7.3587), which achieved 10.8926 and 11.1288 on the Chinese duet dataset, indicating that a model trained with data in the same language will perform better on most songs. Some harmonies are created by the lead vocalist, which is not seen in speech data. The same singer’s voice characteristics are similar, so self-harmonic raises the difficulty of the singer separation task. The experimental results (Table 3) indicate the EN-Duet model performs less well on EN-S, and EN-Self model will outperform the EN-Duet model more than 4\%.

Based on the above experiment, the lyric language and number of singers will affect singer separation results. Applying the model auto-selection method to 14 actual songs taken from the CMedia karaoke APP \footnote{\url{https://www.global-media.com.tw/}}, we obtain the confusion matrix (Table 4). The accuracy rate is 71.43\%, and the average SI-SNRi is 8.9486, which is only 5\% worse than all the correct classification results (9.4611). For the songs for which the EN-Self model scores better, there is no error at all. The above method is feasible for separating vocals in duet and harmonic songs, and this algorithm can also be incorporated into neural network training in the future.

\renewcommand\arraystretch{2} 
\begin{table}[]
\begin{tabular}{cccccc}
 & & \multicolumn{4}{c}{Predicted classes} \\ \cline{3-6} 
 & \multicolumn{1}{c|}{} & \multicolumn{1}{c|}{EN-D} & \multicolumn{1}{c|}{CH-D} & \multicolumn{1}{c|}{EN-S} & \multicolumn{1}{c|}{Total} \\ \cline{2-6} 
\multicolumn{1}{c|}{\multirow{4}{*}{\begin{tabular}[c]{@{}c@{}}Actual \\ classes\end{tabular}}} & \multicolumn{1}{c|}{EN-D} & \multicolumn{1}{c|}{\textbf{2}} & \multicolumn{1}{c|}{1} & \multicolumn{1}{c|}{0} & \multicolumn{1}{c|}{3} \\ \cline{2-6} 
\multicolumn{1}{c|}{} & \multicolumn{1}{c|}{CH-D} & \multicolumn{1}{c|}{3} & \multicolumn{1}{c|}{\textbf{6}} & \multicolumn{1}{c|}{0} & \multicolumn{1}{c|}{9} \\ \cline{2-6} 
\multicolumn{1}{c|}{} & \multicolumn{1}{c|}{EN-S} & \multicolumn{1}{c|}{0} & \multicolumn{1}{c|}{0} & \multicolumn{1}{c|}{\textbf{2}} & \multicolumn{1}{c|}{2} \\ \cline{2-6} 
\multicolumn{1}{c|}{} & \multicolumn{1}{c|}{Total} & \multicolumn{1}{c|}{4} & \multicolumn{1}{c|}{7} & \multicolumn{1}{c|}{3} & \multicolumn{1}{c|}{14} \\ \cline{2-6} 
\end{tabular}
\caption{Confusion matrix of model auto-selection.} 
\label{fig:result2 table}
\end{table}

\section{CONCLUSIONS and FUTURE WORK}
\label{sec:CONCLUSIONS FUTURE WORK}
This paper proposes a singer separation system using music characteristics to separate the song source layer by layer. A single sound source can be successfully separated into an accompaniment component and the individual vocals of two singers. We also created and publicly released four datasets: EN-Duet, CH-Duet, EN-Self, and 3 Channels. We compared the separation of the singing vocal in different languages and for different singers, finding that these factors affect model performance, and thus propose a model auto-selection method to maximize performance. Experimental results show that the proposed method outperforms competing approaches in various scenarios.

Future work will compare songs at different tempos to evaluate the impact on model performance. We will perform data augmentation using a single singer's voice, changing the pitch for the same lyrics for mixing into the ground truth data. After identifying all features that affect singer separation, we will use joint training to replace the singer separation model, producing a more powerful singer separation model.

\bibliographystyle{IEEEbib}
\bibliography{strings}

\end{document}